\documentclass[12pt,preprint]{emulateapj}

\newcommand{\kms}{$\rm {km}~\rm s^{-1}$}

\begin{document}

\title{Very Large Telescope Kinematics for omega Centauri: Further
Support for a Central Black Hole.\footnote{Based on observations
collected at the European Organization for Astronomical Research in
the Southern Hemisphere, Chile (085.D-0928)} }

\shorttitle{Black Hole in $\omega$ Centauri}

\author{Eva Noyola\altaffilmark{1,2},Karl Gebhardt\altaffilmark{3}, Markus Kissler-Patig\altaffilmark{4}, Nora L\"{u}tzgendorf\altaffilmark{4}, Behrang Jalali\altaffilmark{4}, P. Tim de Zeeuw\altaffilmark{4,5}, Holger Baumgardt\altaffilmark{6}} 

\altaffiltext{1}{Max-Planck-Institut f\"ur extraterrestrische Physik, 85748, Garching, Germany} 
\altaffiltext{2}{University Observatory, Ludwig Maximilians University, Munich, D-81679, Germany}
\altaffiltext{3}{Astronomy Department, University of Texas at Austin, Austin, TX 78712, USA} 
\altaffiltext{4}{ESO, Karl-Schwarzschild-Strasse 2 ,85748, Garching, Germany}
\altaffiltext{5}{Sterrewacht Leiden, Leiden University, Postbus
9513, 2300 RA Leiden, The Netherlands}
\altaffiltext{6}{Department of Physics, University of Queensland, Brisbane, QLD 4072, Australia} 
\email{noyola@mpe.mpg.de}

\begin{abstract}
  The Galactic globular cluster $\omega$ Centauri is a prime candidate
  for hosting an intermediate mass black hole. Recent measurements
  lead to contradictory conclusions on this issue. We use VLT-FLAMES
  to obtain new integrated spectra for the central region of $\omega$
  Centauri. We combine these data with existing measurements of the
  radial velocity dispersion profile taking into account a new
  derived center from kinematics and two different centers from the
  literature. The data support previous measurements performed for a
  smaller field of view and show a discrepancy with the results from
  a large proper motion data set. We see a rise in the radial velocity
  dispersion in the central region to 22.8$\pm$1.2 \kms, which
  provides a strong sign for a central black hole. Isotropic dynamical
  models for $\omega$ Centauri imply black hole masses ranging from
  $3.0$ to $5.2\times10^4 M_\odot$ depending on the center. The
  best-fitted mass is $(4.7\pm1.0)\times10^4 M_\odot$.
\end{abstract}


\section{Introduction}\label{intro}

Intermediate-mass black holes (IMBHs) may bridge the gap between
stellar mass black holes and super-massive black holes found in the
center of most galaxies. Their existence is appealing in various ways:
they could extend the $M_\bullet-\sigma$ relation for galaxies
\citep{geb00a,fer00} down to dwarf galaxies and globular clusters, and
present a potential connection to nuclear star clusters
\citep{set10}. They could also be the seeds for super-massive black
holes and alleviate problems with difficulties to account for the
rapid growth necessary to explain massive QSOs at high redshift
\citep{tan09}.

The existence of an IMBH at the center of $\omega$ Centauri (NGC~5139)
has been controversial. \citet{noy08} (hereafter NGB08) obtain
line-of-sight velocity dispersion (LOSVD) measurements using the
Gemini-GMOS integral field unit (IFU). They find a velocity dispersion
rise toward the center implying the presence of a $(4\pm1)\times10^4
M_\odot$ black hole when compared to spherical isotropic dynamical
models. In contrast, \citet{vdm10} (hereafter vdMA10), using proper
motions from {\it HST}-ACS imaging, find a lower black hole mass of
$(1.8\pm0.3)\times10^4 M_\odot$ for an isotropic model and their
profile with a central cusp. Their anisotropic model sets an upper
limit of $7.4\times10^3 M_\odot$. The comparison is complicated by the
fact that the cluster centers between NGB08 and \citet{and10}
(hereafter AvdM10) are separated by $\sim12\arcsec$.

The nature of $\omega$ Centauri has been under discussion for a
while. This object has been regarded as the largest globular cluster
in the Galactic system, but the clear metallicity spread
\citep{nor96,sol05}, as well as a double main sequence
\citep{bed04,pio05} has led to the suggestion that it might be the
stripped core of a dwarf galaxy \citep{fre03,mez05,bek06}. $\omega$
Cen has a large central velocity dispersion of $22\pm4$ \kms\ \citep
{mey95}, as well as a fast global rotation of 8 \kms\ \citep{mer97},
at 11 pc from the center. It is the most flattened Galactic globular
cluster \citep{whi87}, and has a retrograde orbit around the galaxy
\citep{din01}. Using both radial velocities and proper motions
\citet{ven06} calculate a total mass of $2.5\times10^6 M_\odot$,
making $\omega$ Cen the most massive Galactic globular cluster.

The extrapolation of the $M_\bullet-\sigma$ relation for galaxies
\citep{tre02} predicts a $1.3\times10^4 M_{\odot}$ black hole for
$\omega$ Cen. At a distance of $4.8\pm0.3$ kpc \citep{ven06}, the
sphere of influence of such a black hole is $\sim$ 5\arcsec. In this
Letter, we present new VLT-ARGUS data that we compare to previous
measurements.

\begin{figure*}[ht!]
 \epsscale{1.0}
  \plotone{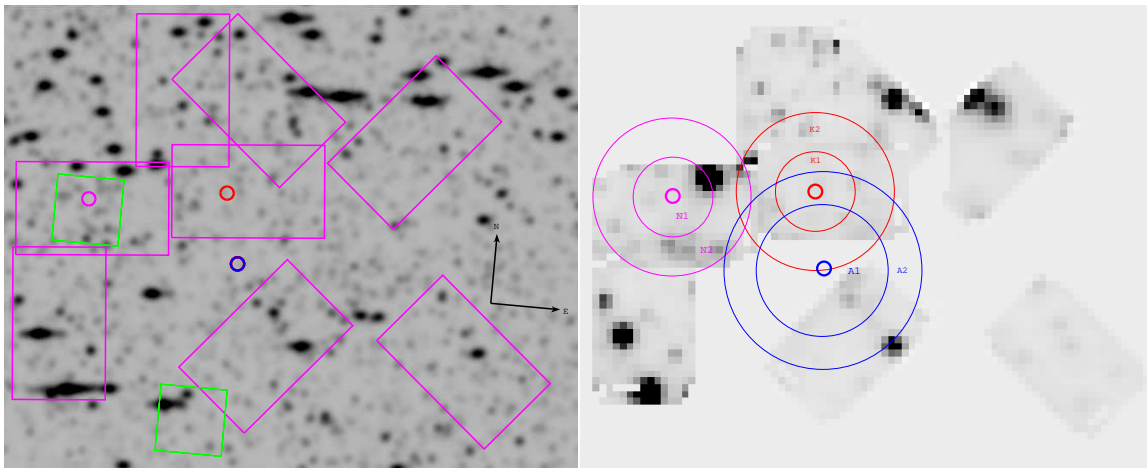}
  \caption{Left: Area of the eight pointings observed with ARGUS
    overlaid on a convolved {\it HST}-ACS image of $\omega$ Centauri
    (about 30\arcsec $\times$ 40\arcsec). The two explored centers
    (magenta circle: NGB08, blue circle: AvdM10), the new kinematic
    center (red), and the previous GMOS IFU pointings (green squares)
    are marked for comparison. Right: reconstructed ARGUS images for
    the eight pointings. The overlaid circles show the two central
    annuli in which the velocity dispersion is measured as a function
    of radius.}
  \label{print}
\end{figure*}

\section{Observations and Data Reduction}

We obtain central kinematics data of $\omega$ Cen using the ARGUS IFU
with FLAMES on the Very Large Telescope (VLT). With a central $\sigma$
around 20 \kms, a spectral resolving power of $R\sim$10,000 is
sufficient to measure the dispersion from integrated stellar
light. The Ca-triplet region (8450--8700\AA) is well suited for
kinematic analysis. The LR8 setup of the GIRAFFE spectrograph
\citep{pas02}, covering the range 820--940 nm at $R\sim$10,400 in
ARGUS mode, is ideally suited for our study.

The ARGUS IFU was used in the 1:1 magnification mode providing a field
of view of 11.5\arcsec$\times$7.3\arcsec, sampled by
0.52\arcsec$\times$0.52\arcsec pixels. The FLAMES observations were
taken during two nights (2009 June 15 and 16). Eight different
pointings were obtained at and around the two contended center
determinations (see Figure \ref{print}). While the pointings aimed at
including both centers, position inaccuracies in the guide star
catalogs made us miss the second from AvdM10. The final set of
observations consists of three exposures for the first ARGUS pointings
(around the NGB08 center) and two exposures for the seven other
pointings, with exposure times of 1500s for the first two, 1020s for
the next two (90$^\circ$ tilted, see Fig. \ref{print}) and 900s for
the four peripheral pointings ($\pm$45$^{\circ}$ tilted).

The first reduction steps are done with the GIRAFFE pipeline (based on
the Base Line Data Reduction Software developed by the Observatoire de
Gen\`eve). The pipeline recipes {\it gimasterbias, gimasterdark,
gimasterflat} and {\it giscience} produce bias corrected, dark
subtracted, fiber-to-fiber transmission and pixel-to-pixel variations
corrected spectra. Sky subtraction and wavelength calibration are
performed with our own tools, which test the wavelength solution with
arc exposures and skylines.

We reconstruct the ARGUS data cubes to images in order to determine
the exact location of the pointings with respect to reference
\textit{Hubble Space Telescope (HST)} images. We use a large Advanced
Camera for Surveys (ACS) mosaic of $\omega$ Cen (GO-9442, PI:
A. Cool), which we convolve to ground-based observed spatial
resolution. The reconstructed ARGUS images are matched to the
convolved ACS image and used to assign the correct location and
position angle from both centers to each pixel and to identify pixels
which are dominated by single stars (i.e., not suited to derive a
velocity dispersion).


\begin{figure}
\epsscale{1.2}
  \plotone{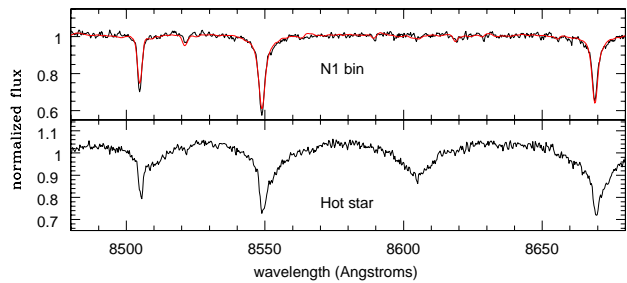}
  \caption{Examples of integrated spectra for the N1 bin (top), as
    well as for one of the hot stars (bottom). The black line shows
    the combined spectra, and the red line shows the fit used for
    kinematics}
  \label{spec}
\end{figure}

\begin{figure*}[ht!]
 \epsscale{1.1}
  \plotone{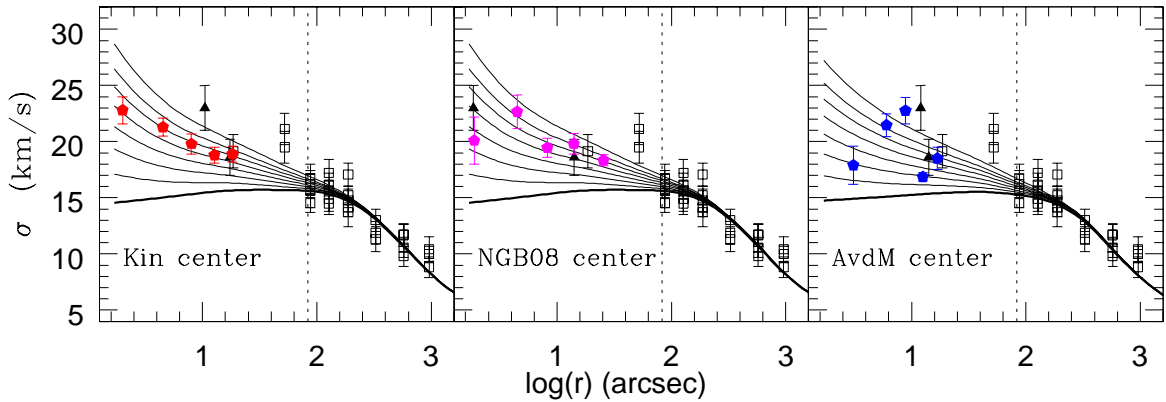}
  \vspace{-8mm}
  \caption{Velocity dispersion as a function of radius. The dashed
    line marks the core radius. The open squares are measurements
    taken from \citet{ven06}. The solid lines show isotropic spherical
    models assuming various black holes masses (0,1,2,3,4,5,6, and
    7.5$\times 10^4~M_\odot$). The left panel shows the measured
    $\sigma$ assuming the new kinematic center (filled red pentagons),
    while the middle and right panels show the same for the NGB08and  AvdM10 center.
    Black triangles mark the kinematic measurements from NGB08.  }
  \label{models}
\end{figure*}


\section{Kinematic Measurements}\label{select}

Measuring kinematics of globular clusters from integral field
spectroscopy is challenging. For details, we refer to NGB08. A key
aspect to consider is the fact that bright stars might dominate the
integrated light and increase the shot noise of the velocity
dispersion . In order to minimize the shot noise from bright stars, we
can choose which pixels to combine for the integrated light measure of
the velocity dispersion.

There are hot stars with strong Paschen-series lines present (see Fig
2). We exclude regions dominated by these stars and those dominated by
bright stars. We identify these regions by including one of these
stars in the velocity template library and then exclude those regions
which have a significant contribution, $\sim$5\% of the pixels are
excluded in this way. We also exclude those regions dominated by
bright stars. After these two cuts, about 85\% of the pixels remain to
derive kinematics. To further minimize the effect from bright stars,
we divide each spectrum by its mean value, thereby giving all pixels
equal weight when combining..

We consider the shot noise from having a small number of stars
contribute to a spatial bin. We calculate the shot noise using the HST
$R$-band photometry from AvdM10. We use Monte Carlo simulations to
generate a mock velocity data set in a given spatial bin, using
magnitudes of present stars. We then estimate a velocity dispersion
weighted by the fluxes of the stars. After 1000 realizations, we get
sample velocity dispersion estimates from which we obtain the scatter,
and hence the shot noise.

We rely on both centers by NGB08 and AvdM10, which differ by
12\arcsec. AvdM10 claim that the center of NGB08 is biased toward
bright stars and that these stars do not trace the center well. On the
other hand, using corrected star counts biases one away from bright
stars. Thus, there may be reasons to expect increased noise for the
center position in both techniques. Given that we have two-dimensional
(2D) kinematics, we can provide another center based on kinematics by
running a kernel of 5\arcsec\ across the field and estimating the
velocity dispersion within that kernel. From the 2D dispersion map
there is a clear peak at the location highlighted in Fig. 2. It lies
about 10\arcsec from NGB08 and 3.5\arcsec\ from AvdM10. We make
dynamical models using the three centers.

We use five annuli centered on each center for the dynamical
analysis. We combine the pixels within each annulus using a biweight
estimator \citep{bee90}. The average radius of the annuli are given in
Table 1, they are chosen to provide a signal-to-noise ratio of at
least 40 in each bin. The central annulus has about 60 pixels and the
outer has 500. The shot noise in any of the outer annuli is below 3\%
of the velocity dispersion. In the central bins, the shot noise is 3\%
for the kinematic center, 6\% for AvdM10 and 9\% for NGB08. The
uncertainties in Table 1 include the shot noise added in quadrature
with the measured uncertainties.

In order to extract the kinematics from the spectra, we use the
technique described in \citet{geb00b} and \citet{pin03}, also employed
in NGB08. This technique provides a non-parametric estimate of the
LOSVD. Starting from velocity bins of 8 \kms, we adjust the height of
each LOSVD bin to define a sample LOSVD. This LOSVD is convolved with
a template. The parameters, bin heights and template mix are changed
to minimize the $\chi^2$ fitted with the data spectrum. For the
template, we use two individual stars within the IFU; these are a
normal late-type giant star, and a hot star (shown in
Fig.~\ref{spec}). The program then determines the relative weight of
these two stars.

\begin{deluxetable}{llllllllll}  
\tabletypesize{\tiny}  
\tablecaption{\label{tab1}Velocity measurements.}
\tablehead{       
\colhead{Bin} &                        
\colhead{$R$} &                        
\colhead{$V$} &                        
\colhead{$\Delta V$} &                      
\colhead{$\sigma$} &      
\colhead{$\Delta\sigma$} &                        
\colhead{$h_3$}  &                     
\colhead{$\Delta h_3$} &
\colhead{$h_4$}  &                     
\colhead{$\Delta h_4$}   \\
\colhead{} &                        
\colhead{\arcsec} &                        
\colhead{$km/s$} &                        
\colhead{$km/s$} &                      
\colhead{$km/s$} &      
\colhead{$km/s$} &                        
\colhead{}  &                     
\colhead{} &
\colhead{}  &                     
\colhead{}  
}                                        
\startdata                               
K1  & 2.0  &  1.1 & 0.5 & 22.8  & 1.2 &  0.05 & 0.03 & -0.05 & 0.01\\
K2  & 4.5  & -1.1 & 0.4 & 21.3  & 0.8 &  0.03 & 0.03 & -0.05 & 0.01\\
K3  & 8.0  &  3.0 & 0.4 & 19.8  & 0.9 &  0.01 & 0.03 & -0.04 & 0.01\\
K4  & 12.7 &  2.3 & 0.4 & 18.8  & 0.7 & -0.01 & 0.02 & -0.04 & 0.01\\
K5  & 18.3 &  1.9 & 0.4 & 18.9  & 0.7 & -0.00 & 0.03 & -0.05 & 0.01\\
N1  & 1.9  & -0.6 & 0.4 & 20.1  & 2.1 &  0.00 & 0.02 & -0.05 & 0.01 \\
N2  & 4.5  &  1.4 & 0.5 & 22.7  & 1.5 & -0.01 & 0.04 & -0.06 & 0.01 \\
N3  & 8.2  &  1.3 & 0.4 & 19.5  & 0.8 & -0.01 & 0.04 & -0.05 & 0.01 \\
N4  & 14.0 &  1.0 & 0.4 & 19.8  & 0.9 &  0.01 & 0.03 & -0.04 & 0.01 \\
N5  & 25.5 &  3.9 & 0.4 & 18.4  & 0.5 &  0.01 & 0.03 & -0.05 & 0.01 \\     
A1  & 3.1  &  8.7 & 0.3 & 17.9  & 1.7 &  0.01 & 0.03 & -0.04 & 0.01\\
A2  & 6.0  & -2.2 & 0.4 & 21.5  & 1.0 &  0.05 & 0.03 & -0.05 & 0.01\\
A3  & 8.8  & -0.5 & 0.5 & 22.8  & 1.2 &  0.00 & 0.03 & -0.08 & 0.01\\
A4  & 12.6 &  4.2 & 0.4 & 16.9  & 0.4 &  0.03 & 0.02 & -0.06 & 0.01\\
A5  & 16.9 &  0.6 & 0.4 & 18.5  & 1.0 & -0.01 & 0.04 & -0.04 & 0.01
\enddata
\tabletypesize{\normalsize}
\end{deluxetable}

The non-parametric LOSVD estimate requires a smoothing parameter (see
\citet{geb00b} for a discussion) in order to produce a realistic
profile, otherwise, adjacent velocity bins can show large
variations. We use the smallest smoothing value just before the noise
in the LOSVD bins becomes large (similar to a cross-validation
technique). In addition to a non-parametric estimate, we fit a
Gaussian--Hermite profile including the first four moments. The second
moment of both the Gauss--Hermite profile and the non-parametric LOSVD
is similar, which implies that we have a good estimate for the
smoothing value. We first fit all individual 4700 pixels in all
dithered positions of the IFU. This step allows us to identify those
pixels where hot stars provide a significant contribution. We then
exclude those pixels from the combined spectra. The top spectrum in
Fig.~\ref{spec} shows the spectral fit to the central radial bin.

The uncertainties for the LOSVD come from Monte Carlo simulations. For
each spectrum, we generate a set of realizations from the best-fitted
spectrum (template convolved with the LOSVD), and add noise according
to the rms of the fit. We then fit a new LOSVD, varying the template
mix. From the run of realizations we take the 68\% confidence band to
determine the LOSVD uncertainties.

Given that $\omega$ Cen contains stars with different spectral types,
we also allow the equivalent widths to be an additional
parameter. This parameter allows for mismatch between the stars chosen
as templates and different regions of the cluster. We have tried a
variety of different template stars and find no significant
changes. Table 1 presents the first four moments ($v$, $\sigma$,
$h_3$, and $h_4$) of a Gauss-Hermite expansion fitted to the
non-parametric LOSVD. We note that the LOSVDs have
statistically-significant non-zero $h_4$ components, which are
important for the dynamical modeling in terms of constraining the
stellar orbital properties.

Fig.~\ref{models} shows velocity dispersions from the LOSVDs for
spectra combined around all centers. Every case shows an increase in
the dispersion compared to data outside 50\arcsec, while isotropic
models without a black hole expect a drop in the central velocity
dispersion. The dispersion profile obtained for the kinematic center
shows a smooth rise, the one for the NGB08 center is still relatively
smooth, while the profile for the AvdM10 center shows larger
variation. While the larger scatter is not evidence that the two
previous centers are not proper, it is suggestive.


\begin{figure}
\epsscale{1.0}
  \plotone{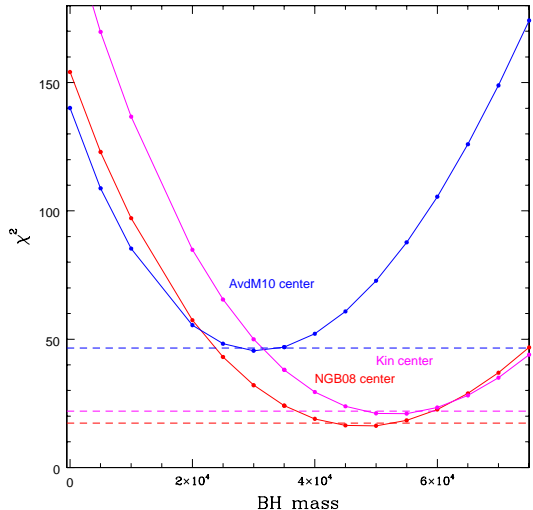}
  \caption{$\chi^2$ for our isotropic, spherical model fits to the
    data, shown for the three cases: the kinematic center (magenta
    curve) the AvdM10 center [blue curve], and the NGB08 ( red
    curve). In both cases the best fit is obtained with a black hole
    of a few $10^4~M_\odot$, while a better fit is achieved assuming
    our original center.}
  \label{chibh}
\end{figure}


\section{Isotropic Models and Discussion}

A detailed comparison with N-body simulations and orbit based models
is in preparation. These models will also consider possible velocity
anisotropy and contribution of dark remnants, as well as include a
comparison with the large proper motion dataset in AvdM10. For the
scope of this Letter, we limit ourselves to a comparison of the
present data with isotropic models. These models have represented the
projected quantities for globular clusters extremely well, starting
with \citet{kin66} all the way to a recent analysis by \citet{mcl06},
where the conclusion is that clusters are isotropic within their
core. Thus, while isotropy needs to be explored in detail, it provides
a very good basis for comparison.

For the details of the isotropic analysis, we refer to NGB08,
essentially following the non-parametric method described in
\citet{geb95b}. The surface brightness profile is the one obtained in
NGB08, which is smoothed and deprojected assuming spherical symmetry
in order to obtain a luminosity density profile \citep{geb96}. By
assuming an $M/L$ ratio, we calculate a mass density profile, from
which the potential and the velocity dispersion can be derived. We
repeat the calculation adding various central point masses ranging
from 0 to $7.5\times10^4M_\odot$ while keeping the global $M/L$ value
fixed. Since vdMA10 obtain a density profile from star counts, we use
their Nuker fit to the star-count profile to create a similar set of
models. We note that the $M/L$ value needed to fit the kinematics
outside the core radius is 2.7 for both profiles. For comparison,
\citet{ven06} found an $M/L$ value of 2.5.

Binaries could potentially bias a velocity dispersion measured from
radial velocities, which is not an issue for proper
motions. \citet{car05} estimate an 18\% binary fraction for $\omega$
Cen (with large uncertainties); this implies that at any given time,
the observed fraction is about a few percent due to chance inclination
and phase \citep{hut92}. Also, \citet{fer06} find no mass segregation
for this cluster tracing the blue straggler population with
radius. Both facts imply that the expected binary contamination is low
(a few percent), which at most would cause a few percent increase in
the measured velocity dispersion (i.e., within our errors).

Figure \ref{models} shows the comparison between the different models
and the measured dispersion profiles. As in our previous study, the
most relevant part of the comparison is the rise inside the core
radius. As can be seen, an isotropic model with no black hole predicts
a slight decline in the velocity dispersion toward the center which is
not observed for any of the assumed centers. The calculated $\chi^2$
values for each model are plotted in Fig.~\ref{chibh}, as well as a
line showing $\Delta \chi^2=1$. The $\chi^2$ curve implies a
best-fitted black hole mass of several $10^4 M_\odot$ in every case,
but with lower $\chi^2$ for the NGB08 center. Specifically, a black
hole of mass of $(5.2\pm0.5)\times10^4M_\odot$ is found for the
kinematic center $(4.75\pm0.75)\times10^4M_\odot$ for the NGB08
center, and of $(3.0\pm0.4)\times10^4M_\odot$ for the AvdM10 center.

The velocity dispersion at 100\arcsec\ is well measured at around 17
\kms. The radial velocities inward show a continual rise in the
dispersion with smaller radii to the central value around 22.8 \kms,
which is statistically significant. This rise is now seen in multiple
radial velocity datasets. It is this gradual rise that provides the
significance for a central black hole. The proper motion data of
AvdM10 show a slight rise in the velocity dispersion, but not all the
way into their center. It is unclear why the two dispersion
measurements differ.

\vspace{-10.0pt}

\acknowledgments 

This research was supported by the DFG cluster of excellence Origin
and Structure of the Universe (www.universe-cluster.de). K.G.
acknowledges support from NSF-0908639. We thank the anonymous referee
for constructive comments that improved the manuscript.

\facility{ESO(VLT-FLAMES)}



\bibliographystyle{apj}

\end{document}